# Nondiffracting vortex-beams in a birefringent chiral crystal

Tatyana A. Fadeyeva and Alexander V. Volyar[*]

*Physics Department, Taurida National V.I. Vernadsky University*
*Vernadsky av.4, Simferopol, Ukraine, 95007*
*e-mail: volyar@crimea.edu*

**Abstract**

*A vector wave analysis of nondiffracting beams propagating along a birefringent chiral crystal for the case of tensor character both of the optical activity and linear birefringence is presented, fields of eigen modes satisfying vector wave equation. We have written characteristic equations and found propagation constants and amplitude parameters of eigen modes. We have shown that the field of eigen modes is non-uniformly polarized in the beam cross-section. We have revealed that even a purely chiral crystal without a linear birefringent can generate optical vortices in an initially vortex-free Bessel beam.*

## I. Introduction

It is well known [1] that nondiffracting beams (or diffraction-free beams) form a wide class of wave fields as solutions to the Helmholtz equation provided that the equation is separable while the beams have transmission symmetry along one of the coordinates (say, the *z*-axis). These solutions refer to different kinds of coordinate systems: Cartesian, circular cylindrical, parabolic cylindrical, and elliptical cylindrical coordinates. The separability of the Helmholtz equation imposes the condition that the solutions of the transverse part not depend on the longitudinal coordinate. The simplest solution in Cartesian coordinates is a plane wave. The most well studied nondifracting wave fields are Bessel beams (see, for example, [2] and references to it) that are the result of the solution to Helmholtz equation in circular cylindrical coordinates. The solution to the Helmholtz equation in elliptic cylindrical coordinates entails Mathieu beams [1] while the employing parabolic cylindrical coordinates results in parabolic diffraction-free beams [3]. The experimental substantiation of the last two kinds of beams have been demonstrated in the papers [4,5]. Notice, to create an ideal nondiffracting beam would require infinite energy. However, in the experiment such a problem is solved by creating approximately nondiffracting beams with the help of special phase plates of finite sizes so that the beam properties do not significantly change over short propagating distances [6, 7]. Improvement of the experimental technique stimulates applications of nondiffracting beams as optical tweezers, in microlithography, metrology, medical imaging, etc [2,8].

One of the key problems of modern singular optics [9] is a control of properties of nondiffracting beams, in particular, optical vortices embedded in them. An appropriate medium for this aim is anisotropic crystals. However, a crystal travelling beam has a vector character. The vector properties of nondiffracting beams, in particular, different types of Bessel beams as solutions to the Maxwell equations in a homogeneous isotropic medium have been analyzed in the papers [10, 11] where the authors showed the beams to be non-uniformly polarized over their cross-section. The analysis of the Bessel beams in biaxial crystals spreading along one of the optical axis was presented in the papers [12,13] where was employed a spectral integral technique. Berry and Jeffray [14] have analyzed the picture of a conical refraction in terms of singular optics focusing their attention on the fact that the Bessel vortex beams in biaxial crystals are the beams with eigen polarization. Notice also that a great number of papers were devoted to properties of paraxial Laguerre-Gaussian and Bessel-Gaussian beams in uniaxial crystals [15-20] where was revealed a unique properties of the crystal travelling beams showing possibilities to generate and annihilate optical vortices.

Analysis of the beam propagation through chiral anisotropic crystals (anisotropic crystals with optical activity) is based as a rule on studying properties of separate plane waves in a beam and expanding them onto the beam as a whole [21]. Berry and Dennis were succeeded in analyzing a fine structure of polarization singularities in a birefringent dichroic chiral crystal on the base of the above approach enhanced by the stereoscopic projection technique [22]. In the paper [23], Berry and Jeffray have analyzed the picture of Poggendorff rings and caustic surface under the condition of conical refraction in a birefringent chiral crystal. The propagation of paraxial Gaussian beam in a birefringent chiral crystal was also studied in [24] on the base of approximate methods. On the other hand, the description of the vortex properties of nondifracting beams in anisotropic chiral crystals on the base of the electromagnetic mode beams technique that reflects the picture of the vortex states in the beam as a whole was not accomplished as yet.

The aim of the given paper is to analyze optical vortex properties in electromagnetic nondiffracting beams traveling through a uniaxial birefringent chiral crystal employing the mode-beam technique.

In Section II, we derive the basic and characteristic equations for nondiffracting electromagnetic beams in a birefringent medium with an anisotropic optical activity. Behavior of a nondiffracting beam in a purely chiral crystal is analyzed in Section III. We solve the characteristic equation for the propagation constants and amplitude parameters of eigen modes and describe evolution of a Bessel singular beam with uniform circular polarization at initial plane of the crystal. Section IV is devoted to solving the characteristic equation in birefringent chiral crystal. Also we analyze optical vortex transformations in a linearly polarized Bessel beam bearing a singly charged optical vortex and vortex-free beams.

## II, The basic and characteristic equations

We consider propagation of a nondiffracting monochromatic beam at the frequency $\omega$ along a unbounded uniaxial anisotropic medium with optical activity (chirality). The Maxwell' equations are:

(a) $\nabla \times \mathbf{E} = -ik\mathbf{B}$, (b) $\nabla \times \mathbf{H} = ik\mathbf{D}$, (c) $\nabla \cdot \mathbf{D} = 0$, (d) $\nabla \cdot \mathbf{B} = 0$ (1)

with a wavenumber $k$ in vacuum while a constitutive equations (see, for example, [25] and references in it) we write in the form:

$$\mathbf{D} = \hat{\varepsilon}\mathbf{E} - i\hat{\gamma}\mathbf{H}, \quad \mathbf{B} = \mathbf{H} + i\hat{\gamma}\mathbf{E}, \quad (2)$$

where the tensors: $\hat{\varepsilon} = diag(\varepsilon, \varepsilon, \varepsilon_3)$ and $\hat{\gamma} = k\hat{g}$, $\hat{g} = diag(g, g, g_3)$ characterize linear birefringence and optically active of a medium, respectively, inherent, for example, in triclinic, hexagonal and other practically important crystal systems, in particular, quartz crystals [26,27]. After simple transformations of the equations (1) and (2) we come to the wave equations for the electric and magnetic fields in the form:

$$\nabla(\nabla \mathbf{E}) - \nabla^2 \mathbf{E} - k^2 \left(\hat{\varepsilon}^2 - \gamma^2 \hat{I}\right)\mathbf{E} = 2k\gamma\nabla \times \mathbf{E} + k\Delta\gamma(\nabla \times \mathbf{e}_z E_z) + \\ + \mathbf{e}_z \Delta\gamma\left[k(\nabla \times \mathbf{E})_z - k^2(\gamma + \gamma_3)E_z\right], \quad (3)$$

$$\mathbf{H} = i(\nabla \times \mathbf{E})/k - i\gamma\mathbf{E} - i\Delta\gamma\mathbf{e}_z E_z \quad (4)$$

with $\hat{I}$ being a unit matrix and $\Delta\gamma = \gamma_3 - \gamma$, $\mathbf{e}_z$ stands for a unit vector of the $z$-axis. On the other hand, from the eqs (1c), (1d) and (2) we derive:

$$\nabla \cdot \mathbf{D} = \left(\varepsilon - \gamma^2\right)\nabla\mathbf{E} + (\Delta\gamma/k)(\nabla \times \partial_z \mathbf{E})_z + \left[\Delta\varepsilon - \Delta\gamma(\gamma + \gamma_3)\right]\partial_z E_z = 0. \quad (5)$$

Besides, using the property of the tensor $\hat{\varepsilon}$ one finds

$$\partial_z E_z = -\left[\left(\varepsilon - \gamma^2\right)/\varepsilon_\gamma\right]\nabla_\perp \mathbf{E}_\perp - \left(\Delta\gamma/k\varepsilon_\gamma\right)(\nabla \times \partial_z \mathbf{E})_z, \quad (6)$$

where $\varepsilon_\gamma = \varepsilon_3 - \gamma_3^2$, $\Delta\varepsilon = \varepsilon_3 - \varepsilon$, $\nabla_\perp = \mathbf{e}_x \partial_x + \mathbf{e}_y \partial_y$, $\mathbf{E}_\perp = \mathbf{e}_x E_x + \mathbf{e}_y E_y$, $\mathbf{e}_x$ and $\mathbf{e}_y$ are unit vectors. So that we obtain

$$\nabla \mathbf{E} = \alpha \nabla_\perp \mathbf{E}_\perp - \left(\Delta\gamma/k\varepsilon_\gamma\right)(\nabla \times \partial_z \mathbf{E})_z, \quad (7)$$

with, $\alpha = \left[\Delta\varepsilon - \Delta\gamma(\gamma + \gamma_3)\right]/\varepsilon_\gamma$. Finally, the basic equation for a birefringent chiral crystal is reduced to the form

$$\nabla^2 \mathbf{E} + k^2\left(\hat{\varepsilon} - \hat{I}\gamma^2\right)\mathbf{E} + 2k\gamma(\nabla \times \mathbf{E}) + \left(\Delta\gamma/k\varepsilon_\gamma\right)\nabla(\nabla \times \partial_z \mathbf{E})_z \\ = \alpha\nabla(\nabla_\perp \mathbf{E}_\perp) - k\Delta\gamma(\nabla \times \mathbf{e}_z E_z) + \mathbf{e}_z k\Delta\gamma\left[k(\gamma + \gamma_3)E_z - (\nabla \times \mathbf{E})_z\right], \quad (8)$$

A longitudinal symmetry of nondiffracting mode beams supposes that

$$\mathbf{E}(x,y,z) = \mathbf{E}(x,y)\exp(-i\beta z), \quad (9)$$

where $\beta$ stands for a propagation constant of the mode beam. As a result, the equations for the transverse and longitudinal components of the electric vector are

$$\nabla_\perp^2 E_x + \bar{U}^2 E_x + 2k\gamma(\nabla \times \mathbf{E})_x + \\ + \Delta\gamma\left[k(\nabla \times \mathbf{e}_z E_z)_x - i(\beta/k\varepsilon_\gamma)\partial_x(\nabla \times \mathbf{E})_z\right] = \partial_x(\nabla_\perp \mathbf{E}_\perp), \quad (10)$$

$$\nabla_\perp^2 E_y + \bar{U}^2 E_y + 2k\gamma(\nabla \times \mathbf{E})_y + \\ + \Delta\gamma\left[k(\nabla \times \mathbf{e}_z E_z)_y - i(\beta/k\varepsilon_\gamma)\partial_y(\nabla \times \mathbf{E})_z\right] = \alpha\partial_y(\nabla_\perp \mathbf{E}_\perp), \quad (11)$$

$$E_z = -i\left(\varepsilon - \gamma^2\right)/\left(\varepsilon_\gamma \beta\right)\nabla_\perp \mathbf{E}_\perp - \Delta\gamma/\left(\varepsilon_\gamma k\right)(\nabla \times \mathbf{E})_z. \quad (12)$$

where $\bar{U}^2 = k^2\left(\varepsilon - \gamma^2\right) - \beta^2$. The above equations can be reduced to a suitable form if one makes use of new variables

$$u = x + iy, v = x - iy, \quad 2\partial_u = \partial_x - i\partial_y, \quad 2\partial_v = \partial_x + i\partial_y, \quad (13)$$

so that

$$\nabla_\perp E_\perp = \partial_v E_+ + \partial_u E_-, \text{ and } \nabla^2 \equiv 4\partial_{uv}^2,$$

where we introduce also a circularly polarized basis:

$$E_+ = E_x - iE_y, \quad E_- = E_x + iE_y, \quad (14)$$

Then the basic equations take a form

$$\nabla_\perp^2 E_+ + \left(\bar{U}^2 - 2k\beta\gamma\right)E_+ - i2(\kappa\beta + \theta)\partial_u(\nabla \times \mathbf{E})_z = 2(\ldots, \quad (15)$$

$$\nabla_\perp^2 E_- + \left(\bar{U}^2 + 2k\beta\gamma\right)E_- - i2(\kappa\beta - \theta)\partial_v(\nabla \times \mathbf{E})_z = 2(\ldots, \quad (16)$$

where, $\theta = \Delta\gamma(\gamma + \gamma_3)/\varepsilon_\gamma$, $\kappa = \Delta\gamma/(k\varepsilon_\gamma)$

$$\Omega = k(\gamma + \gamma_3)(\varepsilon - \gamma^2)/\varepsilon_\gamma,$$
$$(\nabla \times \mathbf{E})_z = i(\partial_v E_+ - \partial_u E_-).$$

We will find a particular solution to the above equations in the form:

$$\mathbf{E} = A\begin{pmatrix}\partial_u \Psi \\ -\partial_v \Psi\end{pmatrix} + B\begin{pmatrix}\partial_u \Psi \\ \partial_v \Psi\end{pmatrix}, \quad (17)$$

where $\Psi = \Psi(x,y)$ is a scalar function, $A, B$ are some parameters. Notice that the first and the second terms in eq.(17) are of a transverse electric (TE) and transverse magnetic (TM) wave fields, respectively, for the case of a pure birefringent crystal $\gamma = \gamma_3 = 0$ [19]. Substituting eq. (17) into eqs (15) and (16) we find the expressions

$$\nabla_\perp^2 \Psi + U_1^2 \Psi = 0, \quad \nabla_\perp^2 \Psi + U_2^2 \Psi = 0, \quad (18)$$

$$U_1^2 = \frac{(\bar{U}^2 - 2k\beta\gamma)(A+B)}{A(1+\kappa\beta+\theta) + B[1+(\Omega/\beta)-\alpha]},$$

$$U_2^2 = -\frac{(\bar{U}^2 + 2k\beta\gamma)(A-B)}{A(1-\kappa\beta+\theta) - B[1-(\Omega/\beta)-\alpha]}. \quad (19)$$

The function $\Psi$ will obey the only equation

$$\nabla_\perp^2 \Psi + U^2 \Psi = 0 \quad (20)$$

provided that $U_1 = U_2 = U$, from whence we come to the expressions

$$[\bar{U}^2 - U^2(1+\theta)]A - [2k\beta\gamma + (\Omega/\beta)U^2]B = 0, \quad (21)$$

$$-\beta(2k\gamma + \kappa U^2)A + [\bar{U}^2 - U^2(1-\alpha)]B = 0. \quad (22)$$

This set of equations has a non-trivial solution relative to amplitude parameters $A$ and $B$ if

$$\begin{vmatrix} k^2(\varepsilon-\gamma^2)-\beta^2-U^2(1+\theta) & -[2k\beta\gamma+(\Omega/\beta)U^2] \\ -\beta(2k\gamma+\kappa U^2) & k^2(\varepsilon-\gamma^2)-\beta^2-U^2(1-\alpha) \end{vmatrix} = 0 \quad (23)$$

The above expression is of a characteristic equation for a propagation constant $\beta$ of nondiffracting beams transmitting through a birefringent chiral crystal. Its solution:

$$\beta_\pm^2 = k^2(\varepsilon+\gamma^2) - \frac{(\varepsilon_3+\varepsilon-2\gamma\gamma_3)}{2\varepsilon_\gamma}U^2 \pm \sqrt{D}, \quad (24)$$

where $D = 4k^4\varepsilon\gamma^2 + 2k^2\gamma(\varepsilon\gamma_3 - \varepsilon_3\gamma + \varepsilon\Delta\gamma)/\varepsilon_\gamma U^2 +$

$+ \left[(\Delta\varepsilon/2)^2 + \Delta\gamma(\varepsilon\gamma_3-\varepsilon_3\gamma)\right]/\varepsilon_\gamma^2 U^4$ together with the condition $A=1$ defines two propagation constants of eigen modes and their amplitude parameters:

$$B_\pm = (\beta_\pm/k)\left[k^2(\varepsilon-\gamma^2) - \beta_\pm^2 - U^2(\varepsilon_3-\gamma^2)/\varepsilon_\gamma\right] \times$$

$$\left[2\beta_\pm^2\gamma + U^2(\varepsilon_3-\gamma^2)(\gamma+\gamma_3)/\varepsilon_\gamma\right]^{-1}. \quad (25)$$

### III. A purely chiral crystal

#### III.1 Characteristic equations

At first we consider the case of a purely chiral crystal without a linear birefringence: $\varepsilon = \varepsilon_3, \Delta\varepsilon = 0$. One distinguishes two ranges of beam parameters: 1) when the parameter $U$ is more smaller than the wave vector $U \ll k\sqrt{\varepsilon}$, and 2) $U$ is comparable with the wave vector $U \sim k\sqrt{\varepsilon}$. The first condition characterizes a so-called paraxial region of the beam propagation while the second one is a so-called non-paraxial region. When $U = 0$, we deal with a $z$-propagating plane wave. Then, the equation (25) gives two values of the propagation constants

$$\beta_\pm = k(n \pm \gamma) \quad (29)$$

while the beam parameters $B_\pm \to \mp 1$. The propagation constant $\beta_-$ describes spreading a right hand polarized (RHP) component $E_+$ of the wave with $B_+ = -1$ whereas the value $\beta_+$ is inherent in the left hand polarized (LHP) component $E_-$ with $B_- = 1$. In the paraxial case of the beam propagation, contribution of the value $\Delta\gamma$ to the propagation process is negligibly small and we obtain from eqs (24) and (25) the beam parameters in the form

$$\beta_\pm = k(n\pm\gamma)\left[1 - (U/k\sqrt{\varepsilon})^2/2\right],$$

$$B_\pm = \mp\left[1 - (U/k\sqrt{\varepsilon})^2/2\right]. \quad (30)$$

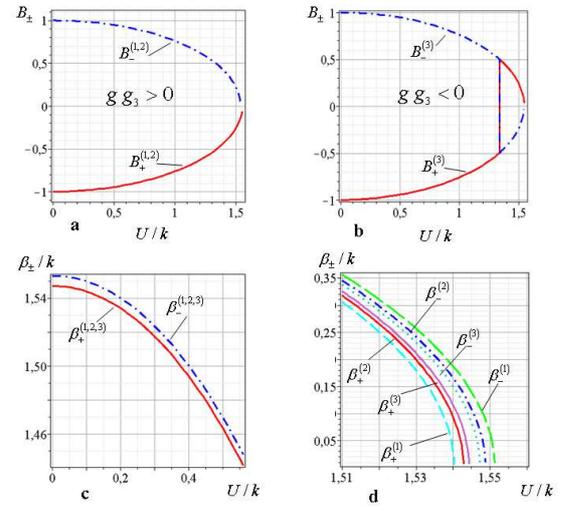

*Fig. 1 (colar on-line) Dispersive curves for the propagation constants $\beta_\pm^{(1,2,3)}/k$ and the mode parameters $B_\pm^{(1,2,3)}$ in a purely chiral medium with $n_o = n_3 = 1.55$: (1)- $g = 3\cdot 10^{-10}$, , $g_3 = 8\cdot 10^{-10}$ (2) - $g = g_3 = 3\cdot 10^{-10}$, (3) - $g = 3\cdot 10^{-10}$, , $g_3 = -8\cdot 10^{-10}$ $[g] = m$*

The tensor character of the optical activity $\hat{g}$ manifests itself the most brightly in the non-paraxial region $U \sim k\sqrt{\varepsilon}$. Indeed, let us require that $D = 0$ in eq. (24). Then we find the characteristic parameter $U = U_{is}$ to be

$$U_{is}^2 = -2k^2(\gamma/\Delta\gamma)\varepsilon_\gamma, \quad \beta_+(U_{is}) = \beta_-(U_{is}). \quad (31)$$

The above equation describes anomaly in a nondiffracting beam-crystal system with isotropic point $U = U_{is}$. Since $\varepsilon \gg \gamma^2$, the wave parameter $U_{is}$ remains a real value only if $\gamma/\Delta\gamma = g/(g_3-g) < 0$. On the other hand, the wave parameter $U_{is}$ cannot exceed the critical value $U_{is} \geq U_{crit} \approx k\sqrt{\varepsilon}$, so that $0 < U_{is}^2 \leq k^2\varepsilon$ or $0 < -g/(g_3 - g) \leq 1$. The last requirement can be presented in a simple form:

$$g_3 g < 0, \qquad |g_3| > |g|. \tag{31a}$$

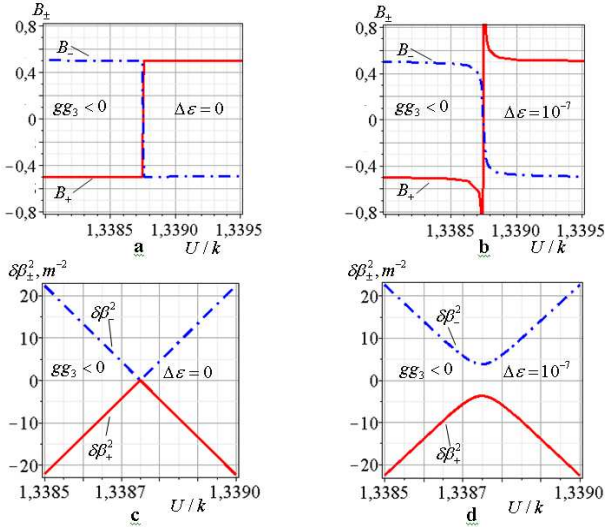

*Fig. 2 (color on-line) Dispersive curves $B_\pm(U)$ and $\delta\beta_\pm^2(U)$ of non-diffracting beams in an optically active medium perturbed by a weak linear birefringence $\Delta\varepsilon$, $g = 3\cdot 10^{-10}$, $g_3 = -8\cdot 10^{-10}$.*

For the rest cases (in particular, for media with isotropic optical activity $\Delta\gamma = 0$), there is not an isotropic point in the beam-crystal system.

Thus, a nondiffracting beam with the U-parameter described by eq. (31) does not feel optical activity. The beam propagates through the chiral crystal as if through an isotropic medium. However, in the vicinity of the isotropic point (31), the beam parameters $B_+$ and $B_-$ change their signs into opposite ones whereas the spectral curves for propagation constants $\beta_+$ and $\beta_-$ experience a sharp bend. *It means that a purely chiral crystal changes handedness of the optical activity to the opposite one with respect to such nondiffracting beams.*

Fig.1a,c illustrates the case without anomaly behavior of nondiffracting beams. The curves in Fig.1c for the propagation constants $\beta_\pm = \beta(U)$ have a smooth form in a broad range of the wave parameters $U$ up to the value $U = U_{crit}$. In the region $U > U_{crit}$ the propagation constant becomes a pure imaginary one and we deal with evanescent nondiffracting beams [28]. The spectral curves $\beta_\pm = \beta(U)$ with the same values of $\gamma$ but different values $\gamma_3$ practically coincide with each other in a paraxial region but have different behavior near critical values $U < U_{crit}$. An anomaly case where the crystal change handedness of its optical activity, is shown in Fig,1b,d. The characteristic curves $B_\pm(U)$ experience sharp changing in vicinity of the isotropic point $U = U_{is}$ that look like a step-transformation both in a large scale of Fig.1b and a small scale of Fig.2a. Such a step-transformation is caused by the fact that the beam parameters are uncertain at the isotropic point $B_\pm(U = U_{is}) = 0/0$. They can have arbitrary values in an isotropic medium. It should notice that the spectral curve $\beta_\pm(U)$ has a sharp band in this range (see Fig.2c). A unique feature of the crystal to change its properties in dependency on the beam parameters forces us to consider behavior of a nondiffracting beam-crystal system rather than properties of the beam and the crystal separately.

Thus, in a purely chiral crystal, the wave field of a simplest nondiffracting beam is characterized by two eigen mode beams (see eq.(17)):

$$\mathbf{E}^{(+)} = \begin{pmatrix} (1+B_+)\partial_u \Psi \\ -(1-B_+)\partial_v \Psi \end{pmatrix} \exp(-i\beta_+ z),$$

$$\mathbf{E}^{(-)} = \begin{pmatrix} (1+B_-)\partial_u \Psi \\ -(1-B_-)\partial_v \Psi \end{pmatrix} \exp(-i\beta_- z). \tag{32}$$

where the upper indices $(\pm)$ in the components of the electric field $\mathbf{E}$ refer to the signs in the propagation constant $\beta_\pm$ and the mode parameters $B_\pm$. The beam field is non-uniformly polarized at the beam cross-

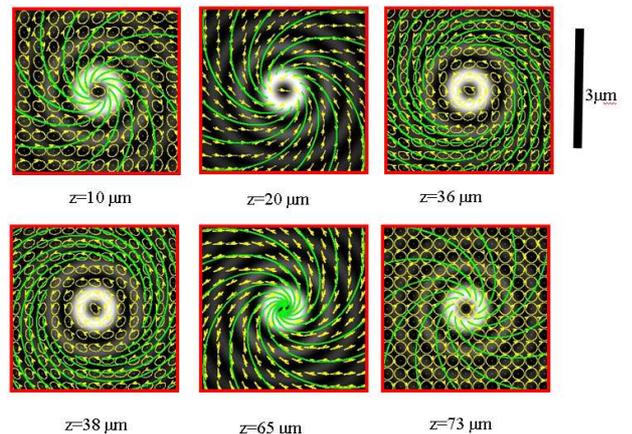

*Fig.3 (color on-line) Maps of the polarization states and integral curves on the background of the intensity distribution: $U = 1.56\cdot 10^7 m^{-1}$, $\Delta g = 0$, $g = 3\cdot 10^{-10}$*

section. In a paraxial region, we can approximately consider that $B_+ \approx -1$ and $B_- \approx 1$ so that the $\mathbf{E}^{(+)}$ field has a dominant LHP component while the $\mathbf{E}^{(-)}$ field has a dominant RHP component. Notice also that the fields

$\mathbf{E}^{(+)}$ and $\mathbf{E}^{(-)}$ have different cut-off parameters $U = U_{\pm}^{(crit)}$ in a non-paraxial region when transforming the propagating eigen modes into the evanescent ones at $\beta_{\pm}(U_{crit}) = 0$.

The absolutely other situation occurs if we will require the field of a nondiffracting beam to be uniformly polarized over all beam cross-section at the plane $z = 0$ in a non-paraxial region. In this case, we have to compensate one of the circular polarizations ($E_+$ or $E_-$) in the superposition of eigen modes in eq. (32). Such a new nondiffracting beam loses its structural stability when transmitting along the crystal. Consider such a process in details on the example of Bessel beams.

### III.2 The Bessel beams

The Helmholtz equation (20) has a solution in circular cylindrical coordinates in the form
$$\Psi_l = J_l(Ur)\exp(\pm il\varphi). \qquad (33)$$
Using equations (33) with $l = 0$ in eq. (32) and taking into account that the operators $\partial_u$ and $\partial_v$ in a polar cylindrical coordinates have a form
$$\partial_u \equiv e^{-i\varphi}\left[\partial_r - (i/r)\partial_\varphi\right],$$
$$\partial_v \equiv e^{i\varphi}\left[\partial_r + (i/r)\partial_\varphi\right]$$
we find the transverse circularly polarized components $\mathbf{E}_{\perp,1}^{(\pm)}$ of the electric field in the form
$$E_{+,1}^{(\pm)} = (1+B_\pm)e^{-i\beta_\pm z}e^{-i\varphi}\left[\partial_r - (i/r)\partial_\varphi\right]\Psi_0,$$
$$E_{-,1}^{(\pm)} = -(1-B_\pm)e^{-i\beta_\pm z}e^{i\varphi}\left[\partial_r + (i/r)\partial_\varphi\right]\Psi_0, \quad (34)$$
where $r^2 = x^2 + y^2$, $\varphi$ stands for an azimuthal angle. The fields $\mathbf{E}_{\perp,1}^{(\pm)}$ are the generatrix vector functions for two major group of fields.

The first group $\mathbf{E}_{|m-1|}^{(\pm)}$ is formed by using a simple relation
$$\mathbf{E}_{|m-1|}^{(\pm)} = \partial_v^m \mathbf{E}_{\perp,1}^{(\pm)} = \left\{e^{i\varphi}\left[\partial_r + (i/r)\partial_\varphi\right]\right\}^{(m)} \mathbf{E}_{\perp,1}^{(\pm)}. \qquad (35)$$
By making use of the Bessel function properties [29]
$$J_{m-1}(x) = J_m'(x) + (m/x)J_m(x),$$
$$J_{m+1}(x) = -J_m'(x) + (m/x)J_m(x)$$
we obtain the first set of Bessel beams:
$$E_{+,|m-1|}^{(\pm)} = (1+B_\pm)e^{-i\beta_\pm z}e^{i(m-1)\varphi}J_{m-1}(Ur),$$
$$E_{-,|m-1|}^{(\pm)} = (1-B_\pm)e^{-i\beta_\pm z}e^{i(m+1)\varphi}J_{m+1}(Ur) \qquad (36)$$
where we omit the factor $(-U)^{m+1}$ in both components and $m = 0,1,2,3,\ldots$. Each circularly polarized component of the beam carries over optical vortex. The vortex in the RHP component has a topological charge $l_+ = m-1$ while the vortex in the LHP component has a topological charge $l_- = m+1$.

The second group of fields $\widehat{\mathbf{E}}_{\perp,m+1}^{(\pm)}$ can be obtained as
$$\widehat{\mathbf{E}}_{\perp,|m+1|}^{(\pm)} = \partial_u^m \mathbf{E}_{\perp,1}^{(\pm)} = \left\{e^{-i\varphi}\left[\partial_r - (i/r)\partial_\varphi\right]\right\}^{(m)} \mathbf{E}_{\perp,1}^{(\pm)} \qquad (37)$$
so that the components of the mode field $\mathbf{E}^{(\pm)}$ take a form:
$$\tilde{E}_{+,|m+1|}^{(\pm)} = (1+B_\pm)e^{-i\beta_\pm z}e^{-i(m+1)\varphi}J_{m+1}(Ur),$$
$$\tilde{E}_{-,|m+1|}^{(\pm)} = (1-B_\pm)e^{-i\beta_\pm z}e^{-i(m-1)\varphi}J_{m-1}(Ur) \qquad (38)$$
The vortices imbedded in the RHP and LHP components of the second group of fields have topological charges $l_+ = -(m+1)$ and $l_- = -(m-1)$, respectively. This situation has much to do with paraxial Laguerre-Gaussian beams in a purely birefringent crystals [16,19].

In the simplest cases, for example, for the beams with uniformly distributed a circular polarization state over the beam cross-section at the $z=0$ plane, the field can be written as a superposition:
$$\mathbf{E}_m = a\mathbf{E}_{|m-1|}^{(+)} + b\mathbf{E}_{|m-1|}^{(-)}. \qquad (39)$$
From this point of view, the Bessel beam in a non-paraxial region $U \sim k\sqrt{\varepsilon}$ without an optical vortex $m = 1$ ($l = 0$ in the initial beam) in the RHP component at the plane $z=0$ transmitting through a purely chiral crystal will generate a doubly charged vortex $l_- = 2$ in the LHP component similar to that for the paraxial Gaussian beam [16,19] in a purely birefringent crystal. Using eqs (36), (39) and the property: $J_{-m}(x) = (-1)^m J_m(x)$, we obtain $a = 1$ and $b = -(1-B_+)/(1-B_-)$. The LHP component gets a simple form
$$E_- = 2i(1-B_-)e^{i\varphi}e^{-i\tilde{\beta}z}\sin(\delta\beta z)J_1(Ur), \qquad (40)$$
where $\delta\beta = (\beta_+ - \beta_-)/2$, $\tilde{\beta} = (\beta_+ + \beta_-)/2$. Each circularly polarized component carries over singly charged optical vortices with opposite handedness of the helices. When propagating along a purely chiral crystal, both the RHP and LHP components are oscillated. At the first glance, these oscillations seem to be strange phenomena. Indeed, a circularly polarized plane wave propagates through a purely chiral crystal without any amplitude transformations. However, we deal with a combined non-paraxial beam here consisting of a great number of circularly polarized plane waves. The wave vectors of the waves lie on a cone surface, projections of their polarization states onto the observation plane containing both the RHP and LHP components. It is this property that is reflected in wave structure of eigen mode beams in eqs (38). The mode state with the only circular polarization at the z=0 plane is not inherent to a non-paraxial beam. One of the components in eq. (39) is suppressed by the interference effect only at the initial plane. As a result, we observe amplitude oscillations of the beam components along the crystal with a period: $\Lambda = 2\pi/\delta\beta$. The amplitude oscillation vanishes in a paraxial region.

Such amplitude oscillations perturb field structure of the beam, a polarization distribution at the beam cross-section being continuously transformed as propagating the beam. Typical maps of the polarization states [19] on the background of the total intensity distribution for a non-paraxial region of the beam propagation are illustrated by Fig.3. The integral spiral-like curves in the figures represent the lines tangential to the big axis of the polarization ellipse at each point of the beam cross-section. There are a great number of singular planes located at the distances $\Delta z = \Lambda$ where the LHP component vanishes. As the beam approaches to these planes, the form of the spiral integral curves starts to be transformed: the spirals either are straightened (see $z = 10\mu m$, $z \approx 65\mu m$ in Fig.3) or tend to a set of concentric circuits (see $z \approx 36\mu m$) to form at the planes $z_n = \Lambda/2(2n+1), n = 0,1,2,3,...$ the fields similar to TE modes with a linear polarization over all beam cross-section. The handedness of the spirals is converted to the opposite one when transiting these planes. However, the handedness of the polarization states is not transformed. Notice that such an effect in gyrotropic crystals has been described in the paper [30] for the energy conversion between TE and TM modes in a Bessel beam on the base of a spectral integral technique.

## IV. A birefringent chiral crystal

### IV.1 A characteristic equation

Singular properties of Bessel beams in a purely birefringent crystal are studied in details in Ref. 11. In this section, we concentrate our attention upon a more general case of a birefringent chiral crystal. The basic properties of the nondiffracting beams in this case are described by the characteristic equations (24) and (25). A general analysis shows that the spectral curves for propagation constants $\beta_\pm$ and beam parameters $B_\pm$ have similar features both for chiral birefringent and purely chiral crystals in a paraxial region because the optical activity suppresses a linear birefringent for the beams travelling along the crystal optical axis (compare Fig.1a and Fig.4a). Major differences in properties of the nondiffracting beam-crystal system appear in a non-paraxial region, a primary role figuring here a tensor character of the optical activity $\hat{\gamma}$. Let us consider the key cases.

1) *The case $g_3 g < 0$, $|g_3| > |g|$*. An isotropic point is unstable one in a beam-crystal system. Even a very small perturbation in the form of a linear birefringence $\Delta\varepsilon \ll \gamma$ deletes the isotropic point $U = U_{is}$ from the characteristic curves $\beta_\pm(U)$ and $B_\pm(U)$. Fig.2b,d shows that the step-transformation in the dependency $B_\pm(U)$ is replaced by smooth curves with a new stable singular point – so-called a gap-point: $U_{is} \to U_g$ while a sharp bend - a point of crossing branches of the spectral curve $\beta_\pm(U)$ is smoothed too – different spectral curves are pushed away. The appearing of the gap-point is evidence of a dominant role of a linear birefringent in the beam-crystal system. In vicinity of this point, a linear birefringence suppresses an optical activity. A form of characteristic curves inside such an anomaly range for the crystals with $\Delta\varepsilon \gg \gamma$ is shown in

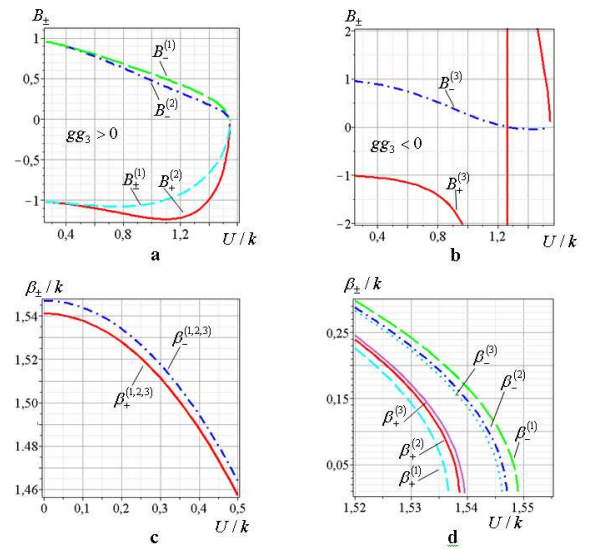

*Fig.4 (color on-line) Dispersive curves for the propagation constants $\beta_\pm^{(1,2,3)}/k$ and the mode parameters $B_\pm^{(1,2,3)}$ in a chiral birefringent crystal with $n_o = 1.544$, $n_3 = 1.553$:  (1)- $g = 3\cdot 10^{-10}$, $g_3 = 8\cdot 10^{-10}$,  (2) -  $g = g_3 = 3\cdot 10^{-10}$,  (3) - $g = 3\cdot 10^{-10}, g_3 = -9\cdot 10^{-10}$*

Fig.4b. The parameter $B_-$ vanishes at the gap-point $U = U_g$: $B_-(U_g) = 0$ whereas the parameter $B_+(U_g) \to -\infty$. A substitution of these conditions in eq. (25) enables us to obtain explicit expressions for the wave parameter $U_g$ and a correction $\Delta\beta^2$ to the propagation constant:

$$U_g^2 = -2k^2(\gamma/\Delta\gamma)\varepsilon_\gamma,$$
$$\Delta\beta^2 = \beta_+^2 - \beta_-^2 = k^2(2\gamma/\Delta\gamma)\Delta\varepsilon. \quad (41)$$

After passing through the gap point, the beam parameters change their signs: $B_- \to -B_-$, $B_+ \to -B_+$ that corresponds to converting handedness of the optical activity to the opposite one (as it takes place in the case of a purely chiral crystal).

2) *The rest cases for the elements of the tensor $\hat{\gamma}$ including $\Delta\gamma = 0$*. Basic features of the beam-crystal system behavior are traced by the dispersive curves in Fig.4a,c. A gap-point disappears in such a system. However, the $B_-$-parameter decreases more quickly than the $|B_+|$-parameter in a non-paraxial range. It means that an optical activity is partially suppressed so that the processes of a linear birefringence stats to control the beam propagation. At the same time, a conversion of the optical activity handedness is lost.

However, the above differences in the beam propagation disappear again near a critical point $U = U_{crit}$ where the propagating beam turns into the evanescent wave.

### IV.2 Singular Bessel beams

High-order Bessel beams carry over optical vortices. Singular properties of scalar vortex-Bessel beams were in details analyzed in [31,32]. We consider here a more general case of transformations of optical vortices in components of the vector Bessel beam with a uniform distributed polarization state at the input $z=0$ of a birefringent chiral crystal.

Let the Bessel beam of the fist order ($m=0$) has a uniform linear polarization at the $z=0$ plane directed along the $x$-axis: $E_x(z=0) = e^{i\varphi}J_1(Ur)$, $E_y(z=0) = 0$. Such a Bessel beam $\mathbf{E}^{(L)}$ can be shaped by four eigen mode beams in eqs (36) with indices $m=0$, $m=2$ and parameters $B_\pm$, $\beta_\pm$:

$$\mathbf{E}^{(L)} = c_1 \mathbf{E}(B_+,\beta_+,m=0) + c_2 \mathbf{E}(B_-,\beta_-,m=0) +$$
$$+ c_3 \mathbf{E}(B_+,\beta_+,m=2) + c_4 \mathbf{E}(B_-,\beta_-,m=2).$$

After simple mathematical transformations, we obtain

$$E_+^{(L)} = \frac{J_1(Ur)e^{-i\bar\beta z}}{2(B_+ - B_-)}\{2i(1+B_-)(1+B_+)\sin(\delta\beta z)e^{-i\varphi} - $$
$$-[(1+B_-)(1-B_+)e^{-i\delta\beta z} - (1-B_-)(1+B_+)e^{i\delta\beta z}]e^{i\varphi}\}, \quad (41)$$

$$E_-^{(L)} = \frac{e^{-i\bar\beta z}}{2(B_+ - B_-)}\{2i(1-B_-)(1-B_+)\sin(\delta\beta z)e^{i3\varphi}J_3(Ur) -$$
$$-[(1+B_-)(1-B_+)e^{i\delta\beta z} - (1-B_-)(1+B_+)e^{-i\delta\beta z}]e^{i\varphi}J_1(Ur)\} \quad (42)$$

It is worth to note that the beam field with the LHP component bearing a positively charged vortex with $l=+1$ at the $z=0$ plane: $E_-(z=0,m=0) \sim e^{i\varphi}J_1(Ur)$ induces in the RHP component a negatively charged vortex $E_+(z,m=0) \sim e^{-i\varphi}J_1(Ur)$ when propagating. On the other hand, the RHP component with a positively charged vortex at the $z=0$ plane $E_+(z=0,m=2) \sim e^{i\varphi}J_1(Ur)$ generates in the LHP component a triply charged vortex: $E_-(z,m=2) \sim e^{i3\varphi}J_3(Ur)$. Thus, the linearly polarized components of the $\mathbf{E}^{(L)}$ beam field in a linearly polarized basis represent combined singular beams bearing single- and triple-charged optical vortices: $E_x = (E_+^{(L)} + E_-^{(L)})/2$, $E_y = i(E_+^{(L)} - E_-^{(L)})/2$. Besides, the eigen modes run with different phase velocities so that the field amplitudes oscillate with a period $\Lambda = 2\pi/\delta\beta$ along the crystal length. Although at the $z=0$ plane, the field is linearly polarized over all cross-section with a centered optical vortex, a little shift of the field along the z-axis results in appearing a complex lattice of optical vortices in each field component. Fig.5 illustrates evolution of the intensity distributions in the $E_x$ component of the vortex-beam with a topological charge $l=1$ (the upper row) in the crystal with $g_3 g < 0$. For comparison, we plotted the evolution of the vortex-free beam ($l=0$) at the initial plane z=0 (the lowest row).

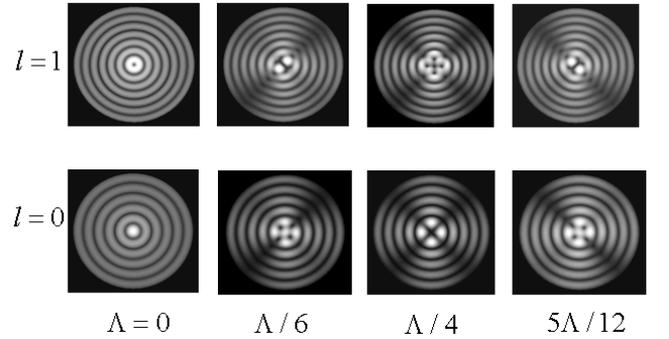

*Fig.5 Intensity distributions in the x-directed linearly polarized components $E_x$ of the vortex Bessel beam ($l=1$) and the zero order ($l=0$) Bessel beam: $g = 3\cdot 10^{-10}, g_3 = -9\cdot 10^{-10}$ $U = 1.3\cdot 10^7 m^{-1}$*

The initial field at the z=0 plane is characterized by a centered optical vortex encircled by a set of ring dislocations. As the beam is slightly shifted along the crystal, a great number of topological dipoles are born along rays: $\varphi = 0$, $\varphi = \pi/2$, $\varphi = \pi$, $\varphi = 3\pi/2$ in the $E_x$ component forming a blur contour of a black cross. At the distance equal to a quarter of a beating length

$L_p = (2p+1)\Lambda/4$, $(p = 0,1,2,...)$, a contrast of the cross becomes more legible. The contrast of the cross becomes maximal at the gap-point $U = U_g$, $z = L_p$ where a a linear birefringence is dominant. Basic difference between the vortex-beams $(l \neq 0)$ at the z=0 and vortex-free beam $(l = 0)$ is that the black cross is shaped as before by optical vortices in the vortex-beams whereas the cross in the vortex-free beam is of two crossed edge dislocations. Any large displacement from the gap-point $U \neq U_g$ results in an essential blurring of the cross contour.

### V. Conclusions

We have derived equations for nondiffracting beams propagating along the optical axis of a uniaxial birefringent chiral crystal with a tensor form both of the linear birefringent and optical activity in constitutive relations and found a characteristic equation for propagation constant and amplitude parameters of eigen modes. We have shown that the eigen modes even in a purely chiral crystals without linear birefringence are non-uniformly polarized at the beam cross-section when a transverse wave number $U$ of Bessel beams is compared with a wave number $k\sqrt{\varepsilon}$ of the electromagnetic beam (so-called a non-paraxial region) while mode beams with $U \ll k\sqrt{\varepsilon}$ (so-called a paraxial region) have nearly uniform polarization distribution. Besides, both circularly polarized components of the Bessel beams carries over optical vortices whose topological charges differ by two units even in a purely chiral crystal. It means that medium with an optical activity can generate optical vortices in an initial vortex-free beam. We have revealed that a tensor character of the optical activity manifests itself the most brightly in the crystals with $g_3 g < 0$, $|g_3| > |g|$. There appears an isotropic point $U = U_{is}$ in a purely chiral crystal when the beam propagates though the chiral crystal as through an isotropic medium. The Bessel beam in a non-paraxial region with uniform circular polarization at the initial plane is accompanied with a periodic recovering of typical conoscopic patterns in a pure chiral.

Even a very slight perturbation with a linear birefringence breaks the isotropic point off the spectral curves of the beam. An optical activity is suppressed by a linear birefringence in this range of the wave parameter $U$. The field structure of the eigen modes in a birefringent chiral crystal is similar to that in a purely chiral crystal but the beam parameters are sufficiently differed. We have revealed that components of a singular Bessel beam with uniform linear polarization in the initial crystal plane have a cross-like conoscopic pattern shaped by a row of optical vortices. The pattern is periodically reconstructed at a quarter of the beam beating length.

### Acknowledgement

The authors is thankful to A.P. Kiselev and R. O. Vlokh for useful considerations of the beam propagation in birefringent chiral crystals.